# Social Network based Short-Term Stock Trading System

Paolo Cremonesi
paolo.cremonesi@polimi.it
Politecnico di Milano, DEIB
Piazza Leonardo da Vinci 32
Milan, Italy

Chiara Francalanci
francala@elet.polimi.it
Politecnico di Milano, DEIB
Piazza Leonardo da Vinci 32
Milan, Italy

Alessandro Poli
poli@elet.polimi.it
Politecnico di Milano, DEIB
Piazza Leonardo da Vinci 32
Milan, Italy

Roberto Pagano
pagano@elet.polimi.it
Politecnico di Milano, DEIB
Piazza Leonardo da Vinci 32
Milan, Italy

Luca Mazzoni
luca.mazzoni@mail.polimi.it
Politecnico di Milano, DEIB
Piazza Leonardo da Vinci 32
Milan, Italy

Alberto Maggioni
alberto3.maggioni@mail.polimi.it
Politecnico di Milano, DEIB
Piazza Leonardo da Vinci 32
Milan, Italy

Mehdi Elahi
mehdi.elahi@polimi.it
Politecnico di Milano, DEIB
Piazza Leonardo da Vinci 32
Milan, Italy

## ABSTRACT
This paper proposes a novel adaptive algorithm for the automated short-term trading of financial instrument. The algorithm adopts a semantic sentiment analysis technique to inspect the Twitter posts and to use them to predict the behaviour of the stock market. Indeed, the algorithm is specifically developed to take advantage of both the sentiment and the past values of a certain financial instrument in order to choose the best investment decision. This allows the algorithm to ensure the maximization of the obtainable profits by trading on the stock market.

We have conducted an investment simulation and compared the performance of our proposed with a well-known benchmark (DJTATO index) and the optimal results, in which an investor knows in advance the future price of a product. The result shows that our approach outperforms the benchmark and achieves the performance score close to the optimal result.

## Categories and Subject Descriptors
H.3.3 [**Information Storage And Retrieval**]: Information Search And Retrieval – *information filtering*

J.1 [**Administrative Data Processing**]: Financial

## 1. INTRODUCTION
This paper focuses on algorithmic equity trading, known as quantitative trading or black-box trading. This indeed is an automated version of equity trading, where algorithms are designed to perform the trading activities by choosing the right time to place the orders along with the number and the price of the stocks to buy or to sell [1]. More specifically, we focus on High Frequency Trading (HFT), which is distinguished by a short-term view. HFT exploits small variations in prices, which are not correlated with long-term investments [2]. Short-term trading represents 60% to 73% of all U.S. equity trading volume [3][4], and 45% of the European equity trading volume [5].

This work starts from the assumption that people's opinions and ratings expressed on social media have an impact on the stock price of the companies providing those products and services.

Since there exists a delay between opinions and financial performance, our goal is to exploit users' buzz to predict the financial market. Although behavioural finance explains that there exist long-term investment waves that are set by large investors rather than people and their mood [7], we wish to verify the idea that smaller investors make short-term investment decisions based on the general mood on an industry or on a specific company.

The main contribution of this paper is three fold:

- a novel predictive trading algorithm that performs sentiment analysis, and adaptively reacts to the market situation by making the appropriate investment decisions.
- a brand model that maps all the important concepts, that may be the subject of Twitter posts, to classes, in order to improve the quality of the sentiment analysis
- an automated trading system that simulates a realistic stock market and allows the (simulated) investor to open either long positions (i.e., buying) or short positions (i.e., selling).

The paper is structured into three parts. In the first part, we develop a brand model in the attempt to enrich a sentiment analysis tool with a domain knowledge specifically targeted mainly on automotive firms and their financial aspects.

In the second part we design an adaptive predictive algorithm that keeps track of past performance measures and tries to constantly change its behaviour in order to better model both sentiment and financial data.

Finally, we run the algorithm to simulate automated investments on the automotive stock market, which enabled us to gain 15% of the initial capital over a three-months period, in comparison to a return of just 3% achieved by the same market, and an optimal return of 20% in the hypothetical scenario in which the investor knows in advance the future price of a product.

## 2. STATE OF THE ART

The belief that market prices can be predicted, at least partially, started from the early critics to the Efficient Market Hypothesis (EMH) [6][7]. EMH is associated with the idea of a random walk, which is typical of a price series where each price is a random result from the preceding prices. From another perspective this is the same of saying that price changes in a given day reflect only the news of that specific day, thus being independent from the preceding price changes. Because of the unpredictability of the news, price changes must be unpredictable and random. This can be summarized as prices fully reflect all known information [7].

The consumers, investors, and managers are somehow driven by different factors which affects their decision making process. One of the most influential factors is the social mood [6]. As behavioural economics provided proofs that financial decisions are significantly driven by emotion and mood [6], in the past years, some researchers have studied if the public mood is correlated or even predictive of economic indicators by taking online social networks as a source of "universal mood" [10][11].

The work presented in [8] focuses on analysis of a discussion forum website[1]. Vector Auto Regression (VAR) analysis revealed that it is no possible to determine whether daily activity on the message board causes or is the result of abnormal returns on the stock. The work in [9] confirmed with significant evidence that messages posted on Yahoo! Finance and Ragingbull website can be used to predict market volatility. The work in [12] describes the usage of Support Vector Machines (SVM) to classify news articles with the purpose of predicting intraday price changes of financial instruments. The system is able to predict whether or not an abnormal return will occur in the next dozen of minutes; however, the direction of returns is not found to be predictable. The sentiment gathered from Yahoo! Finance message board is modelled to be conditionally dependent upon the messages and stock value over the preceding day in [13]. A Naive Bayes model was trained to predict the stock price of Apple and ExxonMobil, with an accuracy ranging from 63% to 81%. In [14] the multiple kernel learning method is used to train a support vector machine with an adaptively-weighted combination of kernels. The resulting model is then applied to predict the stock prices for three Japanese technology companies (i.e., Sharp, Panasonic and Sony) after the extraction of features from financial time series data and from news and comments posted on the Engadget website. Bollen et. al in [10] investigated whether measurements of collective mood states derived from large scale Twitter feeds are correlated to the values of the Dow Jones Industrial Average (DJIA) over time. A Self-Organizing Fuzzy Neural Network (SOFNN) model was then used to predict DJIA values. The model yielded a Mean Absolute Percentage Error (MAPE) of 1.83% and a direction accuracy of 87%.

With respect to the state of the art, a number of key distinguishing aspects, characterizes our work.

First of all, we have designed a brand model in order to map all the important concepts that may be the subject of chatting on Twitter, to drive the sentiment analysis and to better understand the sentiment conveyed by each single post. The brand model has been conceived by following a top-down approach, i.e., including all the aspects that we have considered to be potentially related to price swings of the brands related financial instruments, such as a stock or an index.

The second and the most important distinguishing factor is represented by the model identification approach. As there are no well-accepted methodologies for modelling the sentiment impact on stock markets, we designed an algorithm that could infer some hidden relations from the data in such a way to maximize the profits that could be made by trading the chosen financial instrument. We identified a model over a rather limited time frame, used to predict only the subsequent return, and then repeating this process a number of times by shifting forward the time frame; the algorithm is then in charge of selecting the most proper model by using the various quality metrics defined.

Finally, most of the previous works measure the effectiveness of the trading algorithm by using accuracy metrics (e.g., precision on the direction of the stock price). On the contrary we have assessed the effectiveness of our trading algorithm by performing an extensive trading simulation.

## 3. MODELING THE SOCIAL MOOD

In this section we describe the brand model that forms the core for the sentiment analysis tool.

The first component is a crawler that we used to build our dataset by exploiting the Twitter streaming APIs to collect tweets. All the tweets containing one of the predefined keywords are gathered on a real-time basis.

Keywords are exploited to filter the huge number of tweets that are constantly published on Twitter. The goal is to retrieve all and only those messages that concern some specific brands. The brand represents the subject to be analysed. It can be a single public traded company or an index related to a specific sector, such as automotive or pharmaceutical.

The definition of an effective set of keywords is a typical trade-off problem that requires a balance between precision and recall. Precision is computed as the number of collected messages that are actually related to the brand over the total number of retrieved posts. Recall is computed as the number of collected tweets that are related to the brand over the total number of messages that are theoretically available on the data source. Recall grows as the number of keywords grows; therefore, the idea is to create a sufficiently large set of keywords without affecting precision. The keywords are expressions that their presence in a sentence indicates the relevancy to a specific concept (i.e. the brand). A set of

---

[1] Ragingbull.com

keywords is used to retrieve from the web all the posts that are related to the brand.

We have adopted the following classes of keywords:

- *Brand name:* that is the simplest example of keyword.
- *Names of product:* as a registered trademark.
- *Key people:* CEO, chairman of the supervisory board, etc.

After having cleaned up all the tweets, each post is assigned to a specific brand and further analyzed by mean of a sentiment analysis tool in order to evaluate its polarity (positive, negative or neutral).

For each brand $b$ (i.e., index or stock), the system provides the total amount of positive $P$, negative $N$ and neutral $Z$ tweets with a granularity of half an hour.

Results are then aggregated into a sequence of trading and non-trading sessions. Each session is labelled with the opening and closing prices, as described in Figure 1. First of all, the prices of the selected stocks (i.e. brand) are crawled from Yahoo! Finance and filtered in order to retain the values half an hour after the opening and half an hour before the closing. This is due to the fact that half an hour is considered to be the duration of the price-setting interval for any given stock. These values are then labelled either as for the trading (day) or non-trading (night) sessions. A day value contains, as opening and closing prices, the opening and the closing prices of the corresponding trading session, while a night value contains as the opening price the closing price of the previous trading session and the opening price of the subsequent trading session as the closing price. Weekends are a particular case of non-trading session that lasts for more than one day.

In this way, for each brand $b$ the following 5 variables are obtained, each one describing a time series on the time parameter $t$:

- Financial:

    $O_{bt}$ = opening stocks

    $C_{bt}$ = closing stocks

- Non-Financial:

    $P_{bt}$ = number of positive tweets

    $N_{bt}$ = number of negative tweets

    $Z_{bt}$ = number of neutral tweets

These variables serve as an input to an ad-hoc predictive algorithm that uses both financial and sentiment data.

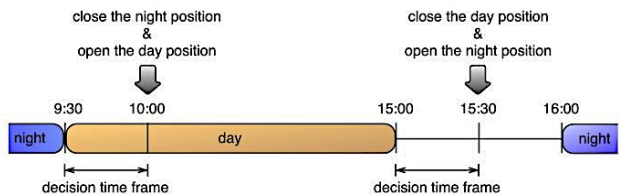

**Figure 1: Definition of trading (day) and non-trading (night) sessions**

## 4. PREDICTING THE STOCK MARKET

The proposed methodology has its core in the identification of different models based on the combination of sentiment and financial variables. What follows is a detailed description of the proposed methodology. The overall conceptual architecture is illustrated in Figure 2, where each component is labelled with a number.

The sentiment variables are obtained by the sentiment analysis component (number 3 in Figure 2). Once a set of Twitter posts is collected (number 2 in Figure 2), they are classified and labelled with a certain sentiment value. The component number 1 downloads the prices of the selected brands from Yahoo, obtains the financial variables! Finance, and computes the corresponding returns. As suggested by most of the financial studies, we considered returns of assets instead of prices [15]. There are several definitions of asset return in the literature. We took the widely used simple net return or simple return $R_t$, which is defined as the following:

*Let $P_t$ be the price of an asset at time index t and lets assume that the asset pays no dividends and that it is held for k periods, from time instant t-k to time instant t. The simple return $R_t$ is*

$$R_t = \frac{P_t - P_{t-k}}{P_{t-k}} \qquad (1)$$

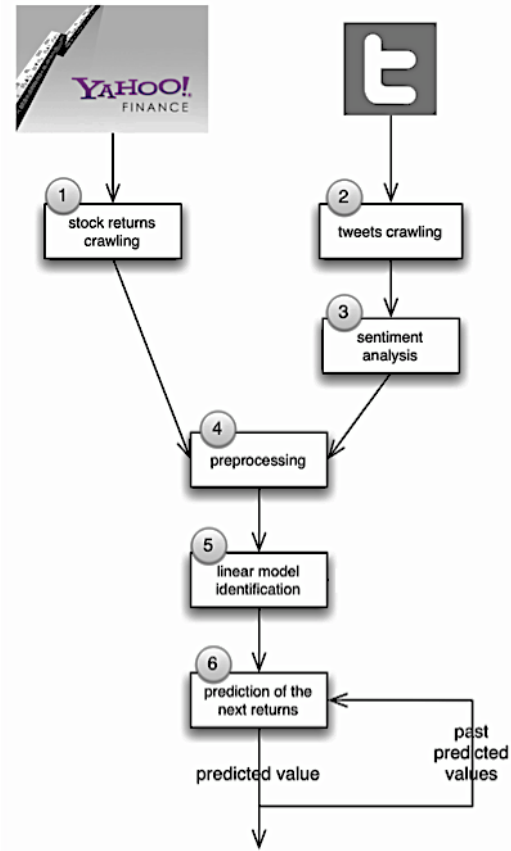

**Figure 2: The overall conceptual architecture**

As stated before, two kinds of time periods have been distinguished, namely day and night (Figure 1). The former is related to the time frame in which the stock market is open and during which the prices of stocks are subjective to variation due to the trading (i.e. the trading session). The latter is related to the time frame during which the stock market is closed and no trading activity can affect the stock prices (i.e. the non-trading session). The previous distinction is essential for the aggregation of the sentiment and for the computation of stocks returns.

A pre-processing stage (number 4 in Figure 2) groups the financial variables together with the sentiment variables in such a way that returns and sentiment values are placed in the correct time point with respect to the previously explained time period.

The model identification stage (number 5 in Figure 2) and the prediction of the next return (number 6) form the core of the system and will be described later on. We postulate that no general model can be conceived in order to explain the stock market phenomena. Hence, for different time frames there could be different combinations of variables that best fit the data. In order to prove that, our system test all the possible combinations of variables for a given period. For each time frame we derive a list of possible models and the system learns how to identify the best model from time to time.

## 4.1 Prediction for a time frame

In this section, we describe the process that chooses the best class of model (sentiment or financial) for a time frame, as summarized in Figure 3.

A fundamental preliminary step is the creation of the power set of all the available variables. This step is where firstly takes shape the idea of an adaptive model that merges the development of a predictive algorithm with the testing phase.

Subsequently, the linear regression model identification is performed for each combination of variables. The underlying idea is that the models that better explain a certain set of recent past values are more likely to correctly predict the next future outcome. Hence, all the models for which all the p-values of their variables are under a fixed threshold of 10% are kept for the forthcoming analysis.

The selected models (i.e., best models in Figure 3) are then classified according to the type of variables they use, i.e. if a model contains solely financial variables, i.e. it uses only the stock returns, is considered to be a financial model, otherwise it is considered to be a sentiment model (so it uses a mix of stock prices and tweets). It is worth nothing that, as financial variables, we selected the pair made up by the 1-lagged and 2-lagged returns. The reason behind this decision is that financial models are purely autoregressive models, which are known to be useful in the analysis of financial time series, while sentiment models have an unknown usefulness until tried. Furthermore, financial models are supposed to behave in a quite different manner with respect to the sentiment ones, because the data series they use are diverse in nature: the former ones manipulate only small numbers that represent percentages, the latter ones a wide variety of data with different meanings. This hypothesis was confirmed by a subsequent analysis of the data, which showed that when financial models occur in conjunction with sentiment models they often provide prediction values that differ in sign and thus are extremely diverse for our purposes.

## 4.2 Financial-Sentiment Spread

Once derived the best models with the p-value filtering, the best model for the current time frame is chosen by considering the history of predictions. We note that we call the number of time frames in the past as Time Frame Window (TFW).

The key aspect of the algorithm is that it is adaptive, i.e., it derives the information from its past performance and exploits it to adapt to changes that may ultimately lead to improvement of the results. The adaptively is embedded in two different levels: (i) the lower level is the choice of the best model between financial and sentiment given a certain TFW, (ii) the higher level is the choice of the best TFW. The underlying idea is that there is a certain degree of correlation in the same series of predictions and that they may provide some interesting and previously unknown cues. Thus, the history of predictions may be usefully employed as a further information source.

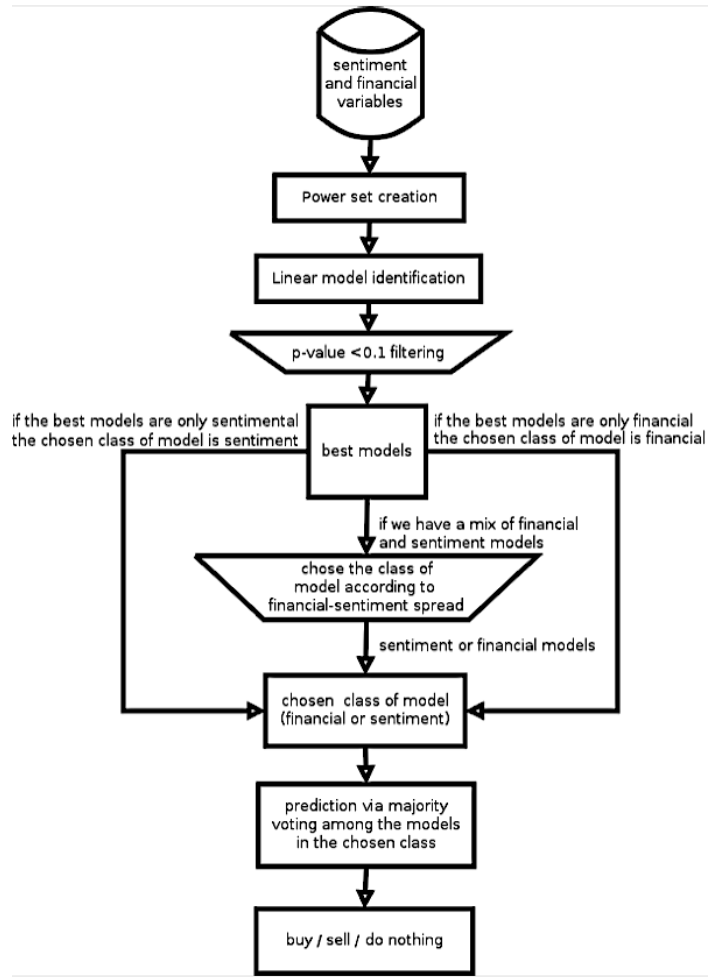

**Figure 3: Activity diagram of the core part of the algorithm**

The variable financial-sentiment spread drives the choice between sentiment and financial models. It is used to keep memory of the relative performance of the former ones with respect to the latter ones, giving an initial advantage to financial models by initially setting it to 1. This is done mainly because sentiment models have to prove their superiority by making more correct predictions with respect to the financial ones as shown in [16] within the life cycle (TFW). The computation of the financial-sentiment spread $s$ for a TF $t$ is calculated, for a real return $r$ as

$$\theta = \begin{cases} +1: & \text{if financial models outperformed} \\ & \text{sentiment models in } TF_{t-1} \\ -1: & \text{otherwise} \end{cases}$$

$$s_t = \gamma\, s_{t-1} + \theta\, |\,100\, r\,| \qquad (2)$$

If the financial-sentiment spread in TF *t-1* is negative, i.e. sentiment models outperform the financial models, the selection of the class of models is made in the same way as if only sentiment models were present in *TF*. Otherwise, financial models are taken as reference for the next prediction. Hence, initially, the algorithm trusts financial models more than the sentiment ones. However, the algorithm is adaptive in the sense that it is able to change the initial decision as soon as changes happen and choose the models that perform better.

The discount factor γ has been introduced in order to assign the correct weight to the history of predictions. Indeed, as our approach was admittedly data-driven, we opted for not fixing it basing on a theoretical reasoning but rather we estimated it by the use of training data.

Now we have to clarify the concept of correctness. For each class of models, the value of correctness measures the percentage of correctly predicted returns (i.e. predicted returns with the same sign of the real ones) that a specific class of models made in a certain TF.

It should be emphasized that the correctness measure is based only on the sign of the predicted return because of a specific reason. We are interested in getting the correct prediction, i.e. a value with the same sign of the next future outcome, because this is the only information needed in order to decide when to buy or sell the stock, and the only one useful. Hence, in this sense, a prediction with the same sign of the real return but quite different in absolute value is more correct than one which is very close to the real return but differs in sign.

Once a single class of models has been chosen, the sign of the return predicted by the majority of the models within that class is then selected to be the predicted trend. This is because at this lowest level of granularity we cannot exploit any other source of information and we are forced to rely on the insight that the majority of a pre-selected list of good models will provide good forecasts. If no model is found to be relevant, the system does not yield any prediction. It is important to underline that for each time frame we choose either a sentiment class or a financial class of models, but not both of them. If no model has been found, the system will not perform any prediction for that time frame ("do nothing" in Figure 3).

## 4.3 The choice of a time frame window

The process just mentioned can be repeated on any number of time frames in the past, thus creating a series of predicted returns that can be compared with the real ones, as shown in Figure 4, in order to simulate an investment scenario.

The process displayed in Figure 4 can be started at an arbitrary time point in the past and can be iterated until it reaches the present time point, thus providing the one step ahead prediction. It is worth noting that, as stated before, this process requires the specification of a TFW.

Hence, at higher level, the same process can be performed with different TFWs, originating different prediction series that can be compared one another. It follows that different TFWs can be used in order to predict the real outcomes at different time points, according to which is the best one at any iteration. In fact, as we cannot know a priori which is the best TFW, the best thing to do is to include for example all the TFWs ranging from 20 time points (i.e. approximately one trading month) to 40 time points, rather than fixing it arbitrarily. Moreover, some TFWs may not produce any relevant model for a specific time point. Therefore, the inclusion of a sufficient number of TFWs helps to overcome such problem so to have one prediction value for every time point.

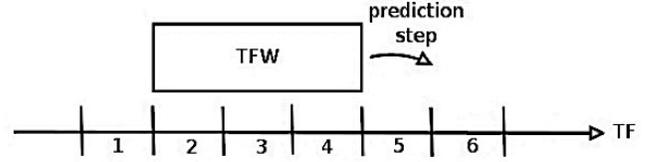

**Figure 4: Scheme that depicts the prediction series resulting from the k times reiteration of the prediction part**

In order to make the comparison, a quality indicator (namely *Q*) associated to each prediction series has been thought. Its calculation for a TF *t* with a real return *r* can be inferred from (3).

$$\lambda_i = \begin{cases} +1: & \text{if the prediction for the TF } i \text{ is correct} \\ 0: & \text{if the TF } i \text{ does not have any prediction} \\ -1: & \text{if the prediction for the TF } i \text{ is wrong} \end{cases}$$

$$Q_t = \beta\, Q\, I_{t-1} + \lambda_i\, |100\, r| \qquad (3)$$

The idea for updating the quality indicator is basically the same as that of the spread. In fact, this level of analysis is higher than that of financial-sentiment spread and thus it seems reasonable to keep the same high-level performance measure based on the investment scenario. Likewise, $\beta$ retains the same meaning of γ but can obviously assume a different value, as the subjects of the comparison differ from one another. Hence *Q* is the indicator at time *t*, $\beta$ is the discount factor, and the absolute value of the real return *r* is the additive component at time *t*.

However, there is a simple but important difference: as most of the TFWs originate prediction series containing values for the majority of the time points, we can state that in this sense their overall behaviour is quite similar. Nevertheless, there are some particular phenomena that should be captured in order to further discount their quality measure.

Basically, we assumed that if there are some missing predictions within a prediction series originated by a given TFW, that is, no relevant models were found, it means that the usefulness of the prediction series created with that TFW for explaining the financial trend in that period has decreased. Hence, there is no reason for having so much trust in it when it will re-appear in the future and its *Q* must be adjusted accordingly. Therefore *Q* is multiplied by $\beta$ at each step, independently from the presence of the prediction value. Moreover, this is another way to increase the reactivity of the algorithm, that is, its ability to rapidly discard those TFWs whose prediction series behaviour is quite unstable.

In order to better clarify the comparison between different prediction series, Figure 5 depicts two prediction series, one originated by the use of TFW $w_1$ and the other originated by the use of TFW $w_n$ where the prediction done by the TF $t_{k-1}$ with a width equal to TFW $w_n$ is chosen due to the fact that the prediction series resulting from the use of TFW $w_n$ has a higher $Q$ with respect to the prediction series resulting from the use of TFW $w_1$.

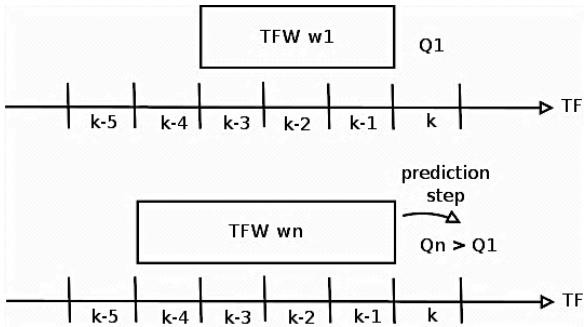

**Figure 5: Comparison between different prediction series**

It must be clear that an arbitrary number of prediction series can be simultaneously compared in order to select the TFW that has originated the prediction series with the highest $Q$. Finally, the chosen prediction is the one made by the majority of the models belonging to the best class, as explained in the previous subsection.

## 5. EXPERIMENTAL RESULTS
In this section we describe the results obtained by the analyses we have conducted.

### 5.1 Analysis A: Quality of the sentiment analysis tool
As the first preliminary experiment, we just tested the quality of the sentiment analysis tool in terms of precision and recall. We examined a sample of more than 1000 posts, performing a manual sentiment analysis. This task is affected by a certain level of subjectivity so that we decided to have three persons independently analyze all the messages, assigning them the most proper brand and a value of sentiment and finally comparing the results until an agreement was reached. The sentiment was classified into positive, negative or neutral, in order to be consistent with the sentiment evaluation component. The overall accuracy level of the brand categorization is 98.6%, while that of the sentiment evaluation is 93%. Other results are reported in Table 1.

**Table 1: Precision and recall levels of the sentiment evaluation**

|  | Positive | Negative | Neutral |
|---|---|---|---|
| **Precision** | 75.81% | 66.67% | 94.58% |
| **Recall** | 51.65% | 55.56% | 97.98% |

### 5.2 Analysis B: Quality of the Investment
The dataset we used contains Twitter posts started from March till September, which is fairly a long period to test the properties of the proposed methodology. The number of tweets was large enough to be statistically significant, as depicted in Figure 6.

In order to perform the tests, we had to choose a proper brand, i.e. an index or a single stock. Even if the algorithm described in the previous section was designed to operate on a daily basis, in principle, it can be also applied using a different scale (e.g. a week, a month, etc.).

However, it is worth noting that not all the time scales are suitable for both types of brand. In particular, a single stock should not be analyzed with a daily granularity, because the volume of posts of a single company, especially of those ones that convey subjectivity, will probably be too low. Therefore, a week seems to be the correct temporal horizon to be adopted. On the contrary, the total number of posts that refers to an index, i.e., a composite of different companies, is equal to the aggregate of all the tweets related to each single company and thus it can be exploited for daily operations.

However, that being considered, we were ultimately forced to select an index as brand due to practical reasons. We crawled the Twitter and performed the analysis on a six-month period. Hence, assuming a week as the reference time unit would have resulted in having approximately 50 points (25 weeks multiplied by 2, because a week is divided as usual into a trading session and the non-trading session of weekends) for training and testing. It stands clear that such a scenario would have been unacceptable and thus we turned to the other option, i.e., an index, in order to be able to use a single day as the temporal reference.

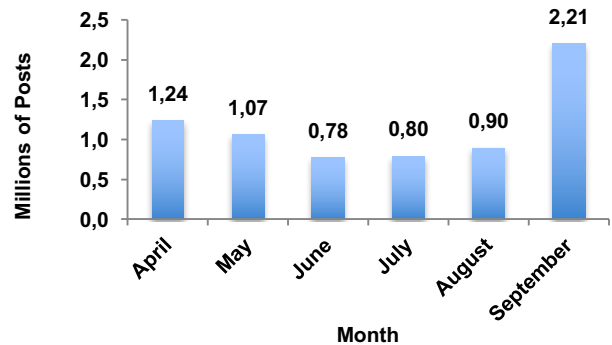

**Figure 6: Monthly posts volume**

For all the reasons exposed above, we elected *Dow Jones Automobiles & Parts Titans 30* index (whose symbol is *DJTATO*) as the index of reference. This index is stated "to represent leading companies in the global Automobiles & Parts sector" and has been regularly computed since 2001 [17]. The wording Titans 30 refers to the fact that its 30 components were selected based on rankings by float-adjusted market capitalization, revenue and net profit [17]. Moreover, it is a U.S. index and thus American investors are expected to be the most relevant ones.

In order to satisfy the 30 minutes time constraint, we developed a parallel version of the algorithm, where a process is spawned for each single TFW, since the work to be done is completely independent from one TFW to the other. In order to give the reader an idea of the time required for the prediction of a single return, the multiprocessing version of the algorithm takes roughly 5 minutes of computation on a quad core processor, which fully satisfies the given time constraint.

As stated before, the parameters $\beta$ and $\gamma$ must be estimated. Hence, we divided the dataset into two different parts, i.e., training and testing. While a sufficient number of points has to be provided for both tasks, we wanted to privilege the testing phase and to avoid any overfitting, and hence, we opted for assigning 30% of the data to the training and 70% to the testing.

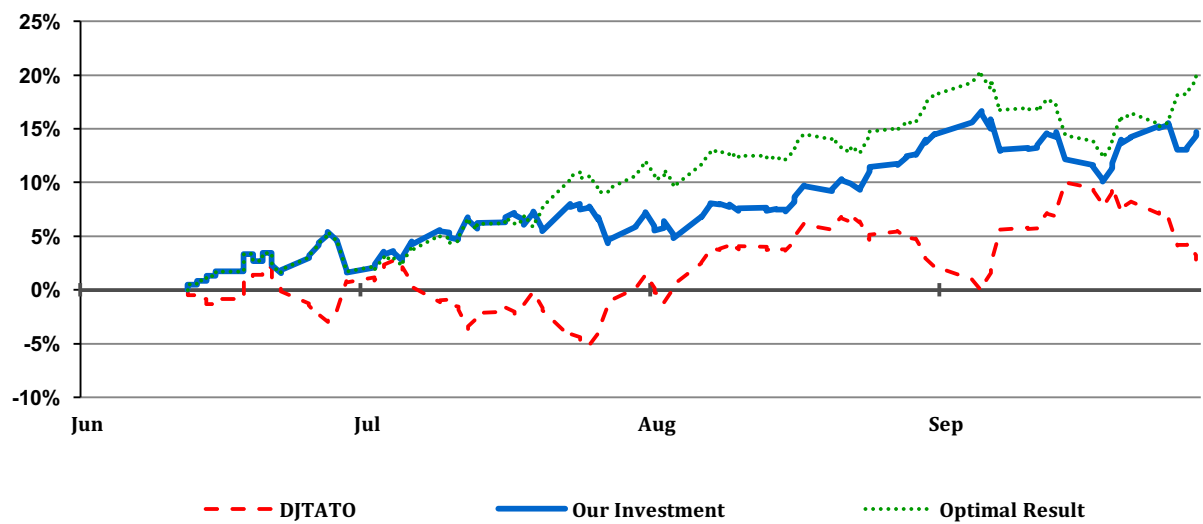

**Figure 7 the result of the analysis**

Moreover, we also faced the problem of defining a proper measure that would have driven the choice of the best values for parameters. In this sense, we agreed that the most appropriate way to test our predictive algorithm was to simulate an investment scenario. Other measures such as Mean Squared Error (MSE), Mean Absolute Error (MAE), etc. seemed to be not suitable for our purposes.

The same holds for training, since we established that the best values for parameters would have been those that maximize the investment return during the training period, which were also expected to provide the largest gain on the overall period. We obtained $\beta = 0.4$ and $\gamma = 0$. It is worth noting that history does not provide any hint in case of financial-sentiment spread ($\gamma$ equal to 0 means that only the last value matters). The same does not hold for $\beta$, i.e. the $Q$ that drives the choice of the most proper TFW by successfully exploiting the history.

Among the other possibilities, we assumed to invest a fixed amount of money at each time step, according to the prediction made by the algorithm. This choice is mainly due to the fact that we preferred to treat each case as independent of the others in order to be able to derive a time-independent performance measure, which does not depend on the particular instant at which the investment has started. Basing on the return, predicted by the algorithm, three different trading decisions become possible:

- *Long:* it refers to the case in which the investor owns the security and thus gains whenever the price goes up. It follows that this option is chosen if the algorithm predicts an increase of DJTATO.

- *Short:* it refers to the sale of stocks that are not owned by the investor. It includes any kind of financial operation that is aimed to make profits from downturns of stocks prices. Hence, it is the option chosen if the algorithm predicts a decrease of DJTATO.

- *No operation:* it is the option chosen if the algorithm does not retrieve any prediction for the current TF.

As explained before, the investment scenario covers two-thirds of the available data, that is, the period that goes from mid-June to the end of September.

**Error! Reference source not found.** illustrates the results of the analysis. As it can be seen our Investment, which represents the cumulative performance obtained by following our proposed approach, clearly outperforms the DJTATO benchmark, which represents the case of an investor holding the DJTATO index, during the entire period of analysis. While the global return of DJTATO is of +3%, interestingly, our investment achieves substantially greater return, i.e., +15%. This is an excellent performance as it is even close to the Optimal Result, i.e., +20%, which represents the best decision-making in which the investor already knows in advance how the market will behave (hypothetical scenario).

This is a promising result as it shows the effectiveness of our proposed algorithm in the prediction of how the market works. Indeed, it clearly indicates the possibility of predicting the behaviour of the market and reacting appropriately in order to maximize the return.

## 6. CONCLUSIONS

With this work, we decided to face an intriguing yet difficult problem, that is the prediction of future stock *returns* by using either the past series of *returns* and the *sentiment* gathered from Twitter.

The quality of the proposed methodology has been thoroughly tested on a reasonable amount of time and it has been shown that the results that we have obtained outperform the benchmark, thus proving the validity of the idea of exploiting the *sentiment* as a fundamental variable.

Even though we tried our best to provide a complete analysis of the stock prediction methodology based on a properly tailored version of the *sentiment* expressed on Twitter, it will be of sure interest to delve into many aspects of the methodology. The first thing will be to apply the same methodology to different sectors such as consumer electronics, telecommunication and others, in order to further validate our approach. Yet another possible development would be the selection of different financial instruments, such as single stocks, derivatives, futures or Exchange-Traded Funds (ETFs).

Another aspect that may deserve attention could be that of augmenting the *sentiment* dimensionality. Moreover, the impact of extending the time lag of the *sentiment* variables could be explored, in such a way to infer something on the possible long lasting effect of the *sentiment*.

Additionally, a study of the nonlinearities could be managed so to discover some possible complex hidden relationships between the *sentiment* and the price deviations.

We strongly believe that the idea of exploiting the *sentiment* conveyed on Twitter for the prediction of price fluctuations of financial instruments is of sure practical interest even though it requires remarkable efforts.